\documentclass[runningheads]{llncs}
\usepackage{calc}
\usepackage{tikz}
\usetikzlibrary{decorations.pathreplacing}
\usepackage{pgfplots}
\usepgfplotslibrary{polar}
\pgfplotsset{compat=newest}
\usetikzlibrary{decorations.markings}
\tikzstyle{vertex}=[circle, draw, inner sep=0pt, minimum size=6pt]
\newcommand{\vertex}{\node[vertex]}
\tikzstyle{vertex}=[circle, draw, inner sep=0pt, minimum size=3pt]
\usepackage{graphicx}
\usepackage{amsmath}
\usepackage{amsfonts}
\usepackage{xcolor}
\usepackage{amssymb}
\usepackage{graphicx}
\usepackage{tikz}
\usetikzlibrary{circuits.logic.US}
\usetikzlibrary{positioning}
\usetikzlibrary{matrix}
\usepackage{graphicx}
\usepackage{amsmath}
\newcommand{\boundellipse}[3]% center, xdim, ydim
{(#1) ellipse (#2 and #3)
}
\definecolor{magenta}{rgb}{0.8, 0.0, 0.8}
\definecolor{cyan}{rgb}{0.0, 1.0, 1.0}
\definecolor{blue1}{rgb}{0.1, 0.6, 0.01}
\definecolor{blue}{rgb}{0.10, 0.50, 1}
\definecolor{brown}{rgb}{0.65, 0.16, 0.16}

\begin{document}

\title{Edge Deletion to Restrict the Size of an Epidemic}
\titlerunning{Edge Deletion to Restrict the Size of an Epidemic}
%\author{ Ajinkya Gaikwad\inst{1} \and Soumen Maity\inst{1} \and  Kitty Meeks\inst{2} 
%\institute{Indian Institute of Science Education and Research, Pune 411008, India 
%&\and
%University of Glasgow, Glasgow G12 8QQ, United Kingdom}\\
%\authorrunning{ S.\,Maity, A.\,Gaikwad}
%\email{\texttt{soumen@iiserpune.ac.in}}
%\email{\texttt{\{ajinkya.gaikwad}}
%}
%
\author{Ajinkya Gaikwad \and Soumen Maity  }
\authorrunning{ A.\,Gaikwad and S.\,Maity}
\institute{Indian Institute of Science Education and Research, Pune, India 
\email{\texttt{ajinkya.gaikwad@students.iiserpune.ac.in;}}
\email{\texttt{soumen@iiserpune.ac.in}}
}
\maketitle              % typeset the header of the contribution
\begin{abstract}

Given a graph $G=(V,E)$, a set $\mathcal{F}$ of forbidden subgraphs, we study {\sc $\mathcal{F}$-Free Edge Deletion}, where
the goal is to remove a minimum number of edges such that the resulting graph does not contain any $F\in \mathcal{F}$ as a
subgraph. For the parameter treewidth,  the question of whether the problem is FPT has remained open.  
%Treewidth is perhaps the most important structural parameter and has been extensively studied for many problems. 
Here we give a negative 
answer by showing that the problem is  W[1]-hard 
when parameterized by the treewidth, which rules out FPT algorithms under common assumption. Thus we give a solution to a conjecture posted by Jessica Enright and Kitty Meeks 
in [Algorithmica 80 (2018) 1857-1889]. We also prove that the {\sc $\mathcal{F}$-Free Edge Deletion} problem is W[2]-hard 
when parameterized by the solution size $k$, feedback vertex set number or pathwidth of the input graph. 
A special case of particular interest is the situation in which $\mathcal{F}$ is the set 
$\mathcal{T}_{h+1}$ of all trees on $h+1$ vertices, so that we delete edges in order to obtain a graph
in which every component contains at most $h$ vertices. This is  desirable from the point of view of restricting the 
spread of disease in transmission network.  We prove that the {\sc $\mathcal{T}_{h+1}$-Free Edge Deletion}
problem is fixed-parameter tractable (FPT) when parameterized by the vertex 
cover number. We also prove that it admits a kernel with $O(hk)$ 
vertices and 
 $O(h^2k)$ edges, when parameterized by combined parameters $h$ and the solution size $k$. 
\end{abstract}

\section{Introduction}
Animal diseases pose a risk to public health and cause damage to businesses and the economy at large. 
%Farmers and the government therefore take every precaution to prevent these diseases, such as keeping animal housing clean and vaccinating livestock.
Among different reasons for livestock disease, livestock movements constitute major routes for the spread of infectious livestock disease \cite{Gibbens729}. For example, the long-range movement of sheep in combination with local transmission resulted in the FMD epidemic in the UK in 2001 \cite{Gibbens729,Mansley43}.  Livestock movements could, therefore, provide insight into the structure of the underlying transmission network and thus allow early detection and more effective management of infectious disease \cite{Danon}.
To do this, mathematical modelling has been employed widely to describe contact patterns of livestock movements and analyse their potential use for designing disease control strategies \cite{Danon}. For the purpose of modelling disease spread among farm 
animals, it is common to consider a  transmission  network with  farms as nodes and 
livestock movement between farms as edges.   

In order to control or limit the spread of disease on this sort of transmission network, we focus our attention on edge
deletion, which might correspond to forbidden trade partners or more reasonably, extra vaccinations or 
disease surveillance along certain trade routes. Introducing extra control of this kind is costly, so it is important to ensure that 
this is done as efficiently as possible. Many properties that might be desirable from the point of view of restricting the 
spread of disease can be expressed in terms of forbidden subgraphs: delete edges so that each connected component 
in the resulting graph has at most $h$ vertices, is equivalent to edge-deletion to a graph avoiding all trees on $h+1$
vertices. We are therefore interested in solving  the following general problem:  
\vspace{3mm}
    \\
    \fbox
    {\begin{minipage}{33.7em}\label{BSP}
       {\sc $\mathcal{F}$-Free  Edge Deletion}\\
        \noindent{\bf Input:} A graph $G=(V,E)$, a set $\mathcal{F}$ of forbidden subgraphs and a positive integer $k$.\\
    \noindent{\bf Question:} Does there exist $E^{\prime}\subseteq E(G)$ with 
    $|E^{\prime}|=k$ such that 
    $G\setminus E^{\prime}$ does not contain any $F\in \mathcal{F}$ as a subgraph?
    \end{minipage} }\\
% \vspace{3mm} \\
 
   \noindent A special case of particular interest is the situation in which $\mathcal{F}$ is the set 
$\mathcal{T}_{h+1}$ of all trees on $h+1$ vertices, so that we delete edges in order to obtain a graph
in which every component contains at most $h$ vertices, so this special case 
    is the problem {\sc $\mathcal{T}_{h+1}$-Free  Edge Deletion}.
\vspace{3mm}
    \\
    \fbox
    {\begin{minipage}{33.7em}\label{BSP1}
       {\sc $\mathcal{T}_{h+1}$-Free  Edge Deletion}\\
        \noindent{\bf Input:} A graph $G=(V,E)$, and two positive integers $k$ and $h$.\\
    \noindent{\bf Question:} Does there exist $E^{\prime}\subseteq E(G)$ with 
    $|E^{\prime}|=k$ such that each connected component in 
    $G\setminus E^{\prime}$  has at most $h$ vertices, that is, 
    the graph $G\setminus E^{\prime}$ does not contain any tree on 
    $h+1$ vertices as a subgraph?
    \end{minipage} }\\
 \vspace{3mm}     
    
 A problem with input size $n$ and parameter $k$ is said to be `fixed-parameter tractable (FPT)' if it has an algorithm that runs in time $\mathcal{O}(f(k)n^c)$, where $f$ is some (usually computable) function, and $c$ is a constant that does not depend on $k$ or $n$. What makes the theory more interesting is a hierarchy of intractable parameterized problem classes above FPT which helps in distinguishing those problems that are not fixed parameter tractable. Closely related to fixed-parameter tractability is the notion of preprocessing. 
    A reduction to a problem kernel, or equivalently, problem kernelization means to apply a data reduction process in polynomial time to an instance $(x, k)$ such that for the reduced instance $(x', k')$ it holds that $(x',k')$ is equivalent to $(x,k)$, $|x'| \leq g(k)$ and $k' \leq g(k)$ for some function $g$ only depending on $k$. Such a reduced instance is called a problem kernel. We refer to~\cite{marekcygan,Downey} for further details on parameterized complexity. 
    
\noindent {\it Our results:} Our main results are the following:
\begin{itemize}
    \item The {\sc $\mathcal{F}$-Free  Edge Deletion} problem is W[1]-hard when parameterized by treewidth.  
   \item The {\sc $\mathcal{F}$-Free  Edge Deletion} problem is W[2]-hard when parameterized by the solution size $k$, the feedback vertex set number or 
   pathwidth of the input graph. 
   .
   \item The {\sc $\mathcal{T}_{h+1}$-Free  Edge Deletion} problem is fixed-parameter tractable (FPT) when parameterized by the
   vertex cover number of the input graph. 
   \item The {\sc $\mathcal{T}_{h+1}$-Free  Edge Deletion} problem  admits a kernel with $O(hk)$ 
vertices and 
 $O(h^2k)$ edges, when parameterized by combined parameters $h$ and the solution size $k$. 
\end{itemize}  

\noindent{\it Previous Work:} If $\pi$ is a graph property, the general edge-deletion problem can be stated as follows: Find the minimum number of edges, whose deletion results in a subgraph satisfying property $\pi$. 
Yannakakis \cite{10.1145/800133.804355} showed that the edge-deletion problem is NP-complete  
 for several common properties, for example,  planar, outer-planar, line-graph, and  transitive digraph.
 Watanabe, Ae, and Nakamra \cite{WATANABE198363} showed that the edge-deletion problem is NP-complete if $\pi$ is
 finitely characterizable by 3-connected graphs.  Natanzon, Shamir and Sharan \cite{NATANZON2001109} proved the NP-hardness of edge-deletion problems with respect to 
 some well-studied classes of graphs. These include perfect, chordal, chain, comparability, split and asteroidal triple free graphs. 
  This problem has also been studied in generality under paradigms like approximation \cite{FUJITO1998213,Lund} and parameterized complexity \cite{CAI1996171,Guo}. FPT algorithms have been obtained for the problem of determining whether there
  are $k$ edges whose deletion results in a split graph \cite{Esha} and to chain, split, threshold, and co-trivially 
  perfect graphs \cite{Guo}. Enright and Meeks \cite{kitty} gave an algorithm for the {\sc $\mathcal{F}$-Free  Edge Deletion} problem 
  with running time $2^{O(|\mathcal{F}|w^r)}n$ where $w$ is the treewidth of the input graph and $r$ is the 
  maximum number of vertices in any element of $\mathcal{F}$. This is a significant improvement on Cai's algorithm
  but does not lead to a practical algorithm for addressing real world problems.
 The special case of this problem in which $\mathcal{F}$
 is the set of all trees on at most $h+1$
vertices is of particular interest from the point of view of the control of disease in livestock, and they have derived an improved algorithm for this special case, running in time $O((wh)^{2w}n)$.

%Given a rooted forest $F$, its \emph{transitive closure} is a graph $H$ in which $V(H)$ contains all the nodes of the rooted forest, and $E(H)$ contain an edge between two vertices only if those two vertices form an ancestor-descendant pair in the forest $F$.
%\begin{definition}
%        {\rm  The {\it treedepth} of a graph $G$ is the minimum height of a rooted forest $F$ whose transitive closure contains the graph $G$. It is denoted by $td(G)$.}
%    \end{definition}

 \section{Hardness of {\sc $\mathcal{F}$-Free Edge Deletion} parameterized by treewidth}  
 In this section we show that {\sc $\mathcal{F}$-Free Edge Deletion} is W[1]-hard parameterized by treewidth, via 
 a reduction from {\sc  Minimum Maximum Outdegree}. Thus we give a solution to a conjecture posted by Jessica Enright and Kitty Meeks \cite{kitty}.\\
    
Let $G=(V,E)$ be an undirected and edge weighted graph, where $V$, $E$, and $w$ denote
 the set of nodes, the set of edges and a positive integral weight 
 function $w:~E\rightarrow Z^{+}$, respectively. An orientation  $\Lambda$ of $G$ is an
 assignment of a direction to each edge $\{u,v\}\in E(G)$, that is, either $(u,v)$ or $(v,u)$
 is contained in $\Lambda$. The weighted outdegree of $u$ on $\Lambda$ is $w_{\mbox{out}}^u=\sum_{(u,v)\in \Lambda}w(\{u,v\})$.
 We define {\sc  Minimum Maximum Outdegree} problem as follows: 
 \vspace{3mm}
    \\
    \fbox
    {\begin{minipage}{33.7em}\label{SP3}
       {\sc  Minimum Maximum Outdegree}\\
        \noindent{\bf Input:} A graph $G$, an edge weighting $w$ of $G$ given in unary, and a positive integer $r$. \\
    \noindent{\bf Question:} Is there an orientation $\Lambda$  of $G$ such that $w_{\mbox{out}}^u\leq r$ for 
    each $u\in V(G)$?
    \end{minipage} }\\
    
\noindent It is known that {\sc  Minimum Maximum Outdegree} is W[1]-hard when parameterized by the treewidth of the input graph \cite{DBLP:journals/corr/abs-1107-1177}.     In this section, we prove the following theorem:

 \begin{theorem}\label{twtheorem}\rm
 The {\sc $\mathcal{F}$-Free Edge Deletion} problem is W[1]-hard when parameterized by the treewidth of the graph.
 \end{theorem}

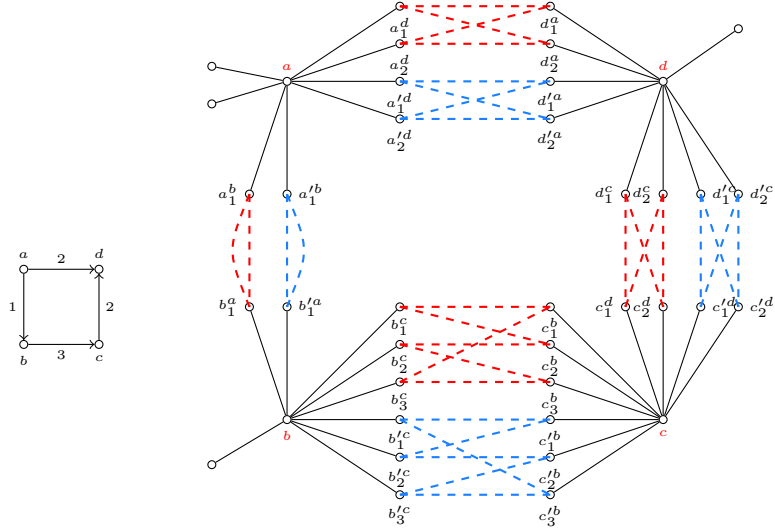
\begin{figure}[ht]
     \centering
    \[\begin{tikzpicture}[scale=1]
    \tiny[square/.style={draw,regular polygon,regular polygon sides=4,minimum size=1},outer sep=0,inner sep=0]
	%\node[square, draw=black] (a1) at (1.5,6.3) [label=above:] {};
    \tikzstyle{triangle}=[draw, shape=regular polygon, regular polygon sides=3,draw,inner sep=0pt,minimum
size=0.25cm]
    %% Notice in the first vertex is named (v) for the sake of a later edge,
	%% and it also has a label to its left that is the math-mode $v$. 
%	\node[triangle, draw=black] (v) at (1,1.5) [label=below:$\triangle^{ab}$] {};  
	%% Notice in the first vertex is named (v) for the sake of a later edge,
	%% and it also has a label to its left that is the math-mode $v$. 
	%\node[vertex, fill=red]
	\vertex (a) at (.5,4.5) [label=above:${\color{red} a}$] {};
%	\node[triangle, draw=black, fill=red] (ha1) at (-0.5,5.2) [label=left:${\color{red} h^{a}_1}$] {}; 
	\vertex (ha2) at (-0.5, 4.7) [] {}; 
	\vertex (ha3) at (-0.5,4.2) [] {}; 
%	\node[triangle, draw=black, fill=red] (ha4) at (-0.5,3.7) [label=left:${\color{red} h^{a}_4}$] {};
	\vertex (ad11) at (2,5.5) [label=below:$a^d_1$] {}; 
	\vertex (ad12) at (2, 5) [label=below:$ a^{d}_2$] {}; 
	\vertex (a'd11) at (2,4.5) [label=below:$a'^d_1$] {}; 
	\vertex (a'd12) at (2,4) [label=below:$a'^{d}_2$] {}; 
    % \vertex (ad21) at (2.5,5.6) [label=below:${\color{blue} a^{\prime d}_1}$] {}; 
%	\node[square, draw=black, fill=blue] (ad22) at (2.5, 4.9) [label=below:${\color{blue} a^{\prime d\square}_1}$] {}; 
%	\vertex (da12) at (4,5.6) [label=below:${\color{red} d^{\prime a}_1}$] {}; 
	\vertex (da11) at (4,5.5) [label=below:$ d^a_1$] {}; 
	%\node[square, draw=black, fill=blue] (da22) at (4,4.9) [label=below:${\color{blue}d^{\prime a \square}_1}$] {}; 
	\vertex (da21) at (4,5) [label=below:$d^{a}_2$] {};
	\vertex (d'a11) at (4,4.5) [label=below:$d'^a_1$] {}; 
	%\node[square, draw=black, fill=blue] (da22) at (4,4.9) [label=below:${\color{blue}d^{\prime a \square}_1}$] {}; 
	\vertex (d'a21) at (4,4) [label=below:$ d'^{a}_2$] {};
	\vertex (b) at (.5,0) [label=below:${\color{red} b}$] {};
%	\node[triangle, draw=black, fill=red] (hb1) at (-0.5,0.6) [label=left:${\color{red} h^{b}_1}$] {}; 
%	\node[triangle, draw=black, fill=red] (hb2) at (-0.5, 0.2) [label=left:${\color{red} h^{b}_2}$] {}; 
%	\node[triangle, draw=black, fill=red] (hb3) at (-0.5,-0.2) [label=left:${\color{red} h^{b}_3}$] {}; 
	\vertex (hb4) at (-0.5,-0.6) [] {};
	\vertex (bc11) at (2,1.5) [label=below:$b^c_1$] {}; 
	\vertex (bc12) at (2,1) [label=below:$b^c_2$] {}; 
	\vertex (bc13) at (2,0.5) [label=below:$b^c_3$] {}; 
	\vertex (b'c11) at (2, 0) [label=below:$ b'^c_1$] {}; 
	\vertex (b'c12) at (2,-0.5) [label=below:$ b'^{c}_2$] {}; 
	\vertex (b'c13) at (2,-1) [label=below:$ b'^{c}_3$] {}; 
     \vertex (ba11) at (0,1.5) [label=left:$ b^a_1$] {}; 
	\vertex (ba13) at (0.5,1.5) [label=right:$b'^{a}_1$] {}; 
%	\vertex (ba21) at (0.4,2) [label=left:${\color{red} b^{\prime a}_1}$] {}; 
%	\node[square, draw=black, fill=blue] (ba23) at (1,2) [label=right:${\color{blue} b^{\prime a \square}_1}$] {}; 
	\vertex (c) at (5.5,0) [label=below:$\color{red} c$] {};
	\vertex (cb11) at (4,1.5) [label=below:$ c^b_1$] {}; 
	\vertex (cb12) at (4, 1) [label=below:$c^b_2$] {}; 
	\vertex (cb13) at (4,0.5) [label=below:$ c^{b}_3$] {}; 
	\vertex (c'b11) at (4,0) [label=below:$c'^{b}_1$] {};
	\vertex (c'b12) at (4,-.5) [label=below:$c'^{b}_2$] {};
	\vertex (c'b13) at (4,-1) [label=below:$c'^{b}_3$] {};
	\vertex (cd11) at (5,1.5) [label=left:$c^d_1$] {}; 
	\vertex (cd12) at (5.5,1.5) [label=left:$ c^d_2$] {}; 
	\vertex (cd13) at (6,1.5) [label=right:$c'^{d}_1$] {}; 
	\vertex (cd14) at (6.5,1.5) [label=right:$c'^{d}_2$] {}; 
	\vertex (d) at (5.5,4.5) [label=above:$\color{red} d$] {};
	\vertex (dc11) at (5,3) [label=left:$ d^c_1$] {}; 
	\vertex (dc12) at (5.5,3) [label=left:$d^c_2$] {}; 
	\vertex (dc13) at (6,3) [label=right:$ d'^{c}_1$] {}; 
	\vertex (dc14) at (6.5,3) [label=right:$d'^{c}_2$] {}; 
	%\vertex (dc21) at (4,3) [label=left:${\color{red} d^{\prime c}_1}$] {}; 
	%\vertex (dc22) at (4.6,3) [label=left:${\color{red} d^{\prime c}_2}$] {}; 
	%\node[square, draw=black, fill=blue] (dc23) at (5.2,3) [label=below:${\color{blue} d^{\prime c \square}_1}$] {}; 
	%\node[square, draw=black, fill=blue] (dc24) at (5.8,3) [label=below:${\color{blue} d^{\prime c \square}_2}$] {}; 
	\vertex (ab11) at (0,3) [label=left:$ a^b_1$] {}; 
	\vertex (ab13) at (.5,3) [label=right:$ a'^{b}_1$] {}; 
	%\vertex (ab21) at (0.4,3) [label=left:${\color{blue} a^{\prime b}_1}$] {}; 
%	\node[square, draw=black, fill=blue] (ab23) at (1,3) [label=right:${\color{blue} a^{\prime b \square}_1}$] {}; 
	\vertex (hd1) at (6.5,5.2) [] {};
%	\node[triangle, draw=black, fill=red] (hd2) at (6.5,4.7) [label=right:$\color{red} h^{d}_2$] {};
%	\node[triangle, draw=black, fill=red] (hd3) at (6.5,4.2) [label=right:$\color{red} h^{d}_3$] {};
%	 \node[triangle, draw=black, fill=red] (hd4) at (6.5,3.7) [label=right:$\color{red} h^{d}_4$] {};
%	\vertex (y) at (1,0.5) [label=below:$\square^{ab}$] {};
	\path
	   % Note that the word "path" here isn't used in the graph-theory sense; the \path command
	   % is always used prior to the list of edges; here, coincidentally, they do form an actual path.
	   %(a) edge (ha1)
	   (a) edge (ha2)
	   (a) edge (ha3)
	   %(a) edge (ha4)
	   (a) edge (ad11)
	   (a) edge (ad12)
	   (a) edge (a'd11)
	   (a) edge (a'd12)
	  % (ad11) edge (ad21)
	   %(ad12) edge (ad22)
	   (d) edge (hd1)
	   (d) edge (dc11)
	   (d) edge (dc12)
	   (d) edge (dc13)
	   (d) edge (dc14)
	   (d) edge (da11)
	   %(da11) edge (da12)
	   (d) edge (da21)
	   (d) edge (d'a11)
	   %(da11) edge (da12)
	   (d) edge (d'a21)
	   %(da21) edge (da22)
	   (a) edge (ab11)
	   (a) edge (ab13)
	   (b) edge (hb4)
	   (b) edge (bc11)
	   (b) edge (bc12)
	   (b) edge (bc13)
	   (b) edge (b'c11)
	   (b) edge (b'c12)
	   (b) edge (b'c13)
       (b) edge (ba11)
	   (b) edge (ba13)
	   (c) edge (cb11)
	   (c) edge (cb12)
	   (c) edge (cb13)
	   (c) edge (c'b11)
	   (c) edge (c'b12)
	   (c) edge (c'b13)
       (c) edge (cd11)
	   (c) edge (cd12)
	   (c) edge (cd13)
	   (c) edge (cd14)
	   ;
       
       %\draw (-2,2) .. controls (-1,0) and (1,0) .. (2,2);
      % \path[draw, dashed] (cb21) -- controls (3,0.5) -- (bc21);
       \draw[thick, dashed,blue] (2,0) -- (4,0);
       \draw[thick, dashed,blue] (2,-.5) -- (4,-.5);
       \draw[thick, dashed,blue] (2,-1) -- (4,-1);
       \draw[thick, dashed,blue] (2,0) -- (4,-1);
       \draw[thick, dashed,blue] (2,-.5) -- (4,0);
       \draw[thick, dashed,blue] (2,-1) -- (4,-0.5);
       \draw[thick, dashed,red] (2,1.5) -- (4,1.5);
       \draw[thick, dashed,red] (2,1) -- (4,1);
       \draw[thick, dashed,red] (2,.5) -- (4,.5);
       \draw[thick, dashed,red] (2,1.5) -- (4,1);
       \draw[thick, dashed,red] (2,1) -- (4,.5);
       \draw[thick, dashed,red] (2,.5) -- (4,1.5);
       \draw[thick, dashed,red] (5,3) -- (5,1.5);
       \draw[thick, dashed,red] (5.5,3) -- (5.5,1.5);
       \draw[thick, dashed,red] (5,3) -- (5.5,1.5);
       \draw[thick, dashed,red] (5.5,3) -- (5,1.5);
        \draw[thick, dashed,blue] (6,3) -- (6,1.5);
       \draw[thick, dashed,blue] (6.5,3) -- (6.5,1.5);
       \draw[thick, dashed,blue] (6,3) -- (6.5,1.5);
       \draw[thick, dashed,blue] (6.5,3) -- (6,1.5);
       \draw[thick, dashed,red] (2,5.5) -- (4,5.5);
       \draw[thick, dashed,red] (2,5) -- (4,5);
       \draw[thick, dashed,red] (2,5.5) -- (4,5);
       \draw[thick, dashed,red] (2,5) -- (4,5.5);
       \draw[thick, dashed,blue] (2,4.5) -- (4,4.5);
       \draw[thick, dashed,blue] (2,4) -- (4,4);
       \draw[thick, dashed,blue] (2,4.5) -- (4,4);
       \draw[thick, dashed,blue] (2,4) -- (4,4.5);
       \draw[thick, dashed,red] (0,3) -- (0,1.5);
       \draw[dashed, thick, red] (0,3) .. controls (-0.3, 2.25) .. (0,1.5);
       \draw[thick, dashed,blue] (.5,3) -- (.5,1.5);
       \draw[dashed, thick, blue] (.5,3) .. controls (.8, 2.25) .. (.5,1.5);
       
      % \draw[dashed] (2.5,-0.3) -- (3.5,-0.3);
      % \draw[dashed] (2.5,-0.3) -- (3.5,0.3);
       %\draw[dashed] (2.5,0.3) .. controls (3.5,0.6) .. (4.5,0.3);
       %\draw[dashed] (2.5,-0.3) .. controls (3.5,0) .. (4.5,-0.3);
      % \draw[dashed] (1.5,0.3) .. controls (2.5,0.6) .. (3.5,0.3);
      % \draw[dashed] (1.5,-0.3) .. controls (2.5,0) .. (3.5,-0.3);
      % \draw[dashed] (4,3) -- (4,2);
      % \draw[dashed] (4.6,3) -- (4.6,2);
      % \draw[dashed] (4,3) -- (4.6,2);
      % \draw[dashed] (2.5,5.6) -- (4,5.6);
      % \draw[dashed] (1.5,5.6) ..controls (2.8,6) .. (4,5.6);
       %\draw[dashed] (2.5,5.6) ..controls (3.55,6) .. (4.6,5.6);
       %\draw[dashed] (4,4) .. controls (3.2,3) .. (4,2);
      % \draw[dashed] (4,3) .. controls (3.2,2) .. (4,1);
      % \draw[dashed] (4.6,4) .. controls (4.9,3) .. (4.6,2);
       %\draw[dashed] (4.6,3) .. controls (4.9,2) .. (4.6,1);
      % \draw[dashed] (0.4,3) -- (0.4,2);
      % \draw[dashed] (0.4,4) .. controls (0.8,3) .. (0.4,2);
      % \draw[dashed] (0.4,3) .. controls (0.8,2) .. (0.4,1);

	\vertex (a) at (-3, 2) [label=above:$a$]{};
	\vertex (d) at (-2,2)  [label=above:$d$]{};
	\vertex (b) at (-3, 1) [label=below:$b$]{};
	\vertex (c) at (-2, 1) [label=below:$c$]{};
	\path[->]
		(c) edge node[right]{$2$}  (d)
		(b) edge node[below]{$3$} (c)
		(a) edge node[left]{$1$} (b)
		(a) edge node[above]{$2$} (d);
		
 \end{tikzpicture}\]
\caption{Result of our reduction on a {\sc Minimum Maximum Outdegree} instance $G$ with $r=2$. The graph $G$ long with its
orientation is shown at the left; and $G^{\prime}$ is shown at the right. 
Complementary vertex pairs are shown using dashed lines. The vertices of the set $V_{\triangle}$ are filled with red color
whereas the vertices of the set $V_{\square}$ are filled with blue color. The vertices in the first part
 of satisfactory partition $(V_1,V_2)$ of $G^{\prime}$
are shown in red label and vertices of $V_2$ are shown in blue label  for the given orientation of $G$. Here $\omega=6$ and $V_0$ contains 64 isolated vertices. 
%The vertices of $V_0$ are
%distributed among $V_1$ and $V_2$ so that $(V_1,V_2)$ becomes balanced satisfactory partition. 
}
     \label{edgedel1}
 \end{figure}

\proof Let  $G=(V,E,w)$ and a positive integer $r \geq 3$ be an  instance $I$ of 
{\sc Minimum Maximum Outdegree}. We construct an instance $I'=(G',k,\mathcal{F})$ 
of {\sc $\mathcal{F}$-Free Edge Deletion} the following way. See Figure \ref{edgedel1} for an illustration.
For each edge 
$(u,v)\in E(G)$, we introduce the following sets of new vertices $V_{uv}=\{u^v_1, \ldots, u^v_{w(u,v)}\}$, $V^{\prime}_{uv}=\{u^{\prime v}_1, \ldots, u^{\prime v}_{w(u,v)}\}$
$V_{vu}=\{v^u_1, \ldots, v^u_{w(u,v)}\}$ and $V^{\prime}_{vu}=\{v^{\prime u}_1, \ldots, v^{\prime u}_{w(u,v)}\}$. 
We make $u$ (resp. $v$) adjacent to all the 
vertices in $V_{uv}\cup V'_{uv}$ (resp. $V_{vu}\cup V'_{vu}$). 
Let $E_{u,u^v}=\Big\{ (u,x) ~|~  x\in V_{uv}\Big\}$, 
$E^{\prime}_{u,u^{\prime v}}=\Big\{ (u,x) ~|~  x\in V^{\prime}_{uv}\Big\}$, 
$E_{v,v^u}=\Big\{ (v,x) ~|~  x\in V_{vu}\Big\}$ and $E^{\prime}_{v,v^{\prime u}}=\Big\{ (v,x) ~|~  x\in V^{\prime}_{vu}\Big\}$. 
Let $\omega =\sum\limits_{e\in E}{w(e)}$
and $N=n+3\omega +1$. 
 Let $\delta(v;E)$ denote the set of edges
in $E$ incident to $v\in V$. The weighted degree $d_w(v;G)$ of a vertex $v\in V$ is 
defined as $\sum\limits_{e\in \delta(v,E)}{w(e)}$. The weighted maximum degree 
$\Delta_w(G)$ of $G$ is defined as 
$\max\limits_{v\in V} d_w(v;G)$.
For every vertex $u\in V(G)$, we also add a set 
$V_u^{\square}$ of $\Delta_w(G)-d_w(v;G)$ many one degree vertices. We define two sets of pair of vertices:
\begin{align*}
 C_1 &=\Big\{\{(u_i^{\prime v},v_i^{\prime u})\}~|~ (u,v)\in E(G), 1\leq i\leq w(u,v)\Big\}\\
& \bigcup \Big\{(u_{i}^{\prime v},v_{i+1}^{\prime u}), (u_{w(u,v)}^{\prime v} , v_{1}^{\prime u}) ~|~ (u,v)\in E(G), 1\leq i\leq w(u,v)-1\Big\}, 
\end{align*} 
\begin{align*}
 C_2 &=\Big\{\{(u_i^{v},v_i^{u})\}~|~ (u,v)\in E(G), 1\leq i\leq w(u,v)\Big\}\\
& \bigcup \Big\{(u_{i}^{v},v_{i+1}^{u}), (u_{w(u,v)}^{v} , v_{1}^{u}) ~|~ (u,v)\in E(G), 1\leq i\leq w(u,v)-1\Big\}.   
\end{align*}
For every pair of vertices $(u^{\prime v},v^{\prime u})\in C_1$, we add a $4N-2$ length blue path 
$P^{\prime}_{(u^{\prime v},v^{\prime u})}$ joining $u^{\prime v}$ and $v^{\prime u}$, whose internal vertices were not originally part  of $G$. Similarly, for every pair of vertices 
$(u^{v},v^{u})\in C_2$, we add an $N$ length red path 
$P_{(u^{v},v^{u})}$  joining $u^{v}$ and $v^{u}$, whose internal vertices were not originally part  of $G$. 
Now, we define the unweighted graph $G'$ as follows:
\begin{align*}
V(G') &= V(G) \bigcup\limits_{u\in V(G)} V_u^{\square}
\bigcup\limits_{(u,v)\in E(G)} (V_{uv} \cup V'_{uv} \cup V_{vu} \cup V'_{vu} )\\&
\bigcup\limits_{(u^{\prime v},v^{\prime u})\in C_1} V(P^{\prime}_{(u^{\prime v},v^{\prime u})})
\bigcup\limits_{(u^{v},v^{u})\in C_2} V(P_{(u^{v},v^{u})})
\end{align*} 
and
\begin{align*}
    E(G^{\prime})&= \bigcup\limits_{u\in V(G)} \{(u,\alpha) \ | \ \alpha \in V_u^{\square}\}
    \bigcup\limits_{(u,v)\in E(G)} {E_{u,u^v}} \cup {E_{v,v^u}} \cup E^{\prime}_{u,u^{\prime v}} \cup E^{\prime}_{v,v^{\prime u}} \\
   & \bigcup\limits_{(u^{\prime v},v^{\prime u})\in C_1} E(P^{\prime}_{(u^{\prime v},v^{\prime u})})
\bigcup\limits_{(u^{v},v^{u})\in C_2} E(P_{(u^{v},v^{u})})
\end{align*} where $V(P)$ and $E(P)$ denote the set of vertices and edges 
of $P$ respectively.
\noindent We set $k=\omega$ and $\mathcal{F}=\{S_{\Delta_w(G) + r+1},C_{5N+2}\}$ where $S_{\Delta_w(G) +r+1}$ is the star graph or
the complete bipartite graph $K_{1,\Delta_w(G) +r+1}$ and $C_{5N+2}$ is the cycle of length $5N+2$. 
%Here, we observe that the degree of $x\in V(G)$ in $G^{\prime}$ is exactly $\Delta_w(G) + d_w(x;G)$. 
We observe that the gadget replacing every edge $(u,v)\in E(G)$ has treewidth at most eleven because 
deleting the set $$\Big\{u,u_{1}^{v}, u_{w(u,v)}^{v}, u_{1}^{\prime v}, u_{w(u,v)}^{\prime v}, v, v_{1}^{u}, v_{w(u,v)}^{u}, v_{1}^{\prime u}, v_{w(u,v)}^{\prime u}\Big\}$$ of vertices makes it a forest. This implies that the treewidth of 
$G'$ is a at most treewidth of $G$ plus eleven.   Now we show that our reduction is correct. That is, we prove that $(G,w,r)$ is  a yes instance of 
{\sc Minimum Maximum Outdegree} if and only if $I^{\prime}$ is a yes instance of {\sc $\mathcal{F}$-Free Edge Deletion}.\\

Let $D$ be the directed graph obtained by  an orientation of the edges of $G$ such that for 
each vertex the sum of the weights of outgoing edges is at most $r$. 
%Let $w_{\mbox{out}}^x$ and $w_{\mbox{in}}^x$ denote the sum of the weights of outgoing and incoming edges of vertex $x$, respectively. 
We claim that the set of edges $$E^{\prime}= \bigcup_{(u,v)\in E(D)} \big\{(x,v) \ | \ x \in V_{vu} \big\}$$ is a 
solution of $I^{\prime}$.
Clearly, we have $|E^{\prime}|=k$. 
 We need to show that, $\widetilde{G^{\prime}}=G^{\prime} \setminus E^{\prime}$  does not contain any forbidden graph
 as a subgraph. 
 First we show that every vertex has degree at most $\Delta_w(G) +r$ in $\widetilde{G^{\prime}}$. 
 It is clear from construction that if $x\in V(G')\backslash V(G)$ then $d_{\widetilde{G^{\prime}}}(x)\leq 3$. 
 %We also observe that for every vertex $x\in V(G)$, every outgoing edge $(x,u)$ contributes exactly $|V_{xu}|=\omega(x,u)$ many edges adjacent to $x$ as shown in figure.
 Let $w_{\mbox{out}}^x$ and $w_{\mbox{in}}^x$ denote the sum of the weights of outgoing and incoming edges of vertex $x$, respectively. Note that $d_w(x;G)=w_{\mbox{out}}^x+w_{\mbox{in}}^x$ and $x$ is adjacent to 
 $\Delta_w(G)-d_w(x;G)+w_{\mbox{in}}^x+2w_{\mbox{out}}^x$ many vertices in $\widetilde{G^{\prime}}$.
 This implies that $d_{\widetilde{G^{\prime}}}(x)\leq \Delta_w(G) +r$ as $w_{\mbox{out}}^x\leq r$. Therefore, 
 $\widetilde{G^{\prime}}$ does not contain $S_{\Delta_w(G) +r+1}$ as a subgraph. Next, 
 we prove that $\widetilde{G^{\prime}}$ does not contain $C_{5N+2}$ as a subgraph. 
 Suppose, for the sake of contradiction, $\widetilde{G^{\prime}}$ contains $C_{5N+2}$ as a 
 subgraph. We make two cases based on whether the cycle contains some original vertex $u$ from $ V(G)$ or not.  \\
 \noindent{\it\bf Case 1:} Let us assume that the cycle includes at least one original vertex $u \in V(G)$. 
Further, we make two subcases based on whether the cycle contains a blue edge or not. \\ 
{\it Subcase 1.1:}  Let us assume that the cycle includes 
at least one blue edge from a blue
path $P^{\prime}_{(u^{\prime v},v^{\prime u})}$. 
Then the cycle  includes all the blue edges of 
$P^{\prime}_{(u^{\prime v},v^{\prime u})}$ and reaches the vertex $v^{\prime u}$.
Without loss of generality, we assume that the direction of edge $\{u,v\}$ is from $u$ to $v$
in $D$. Then the edges in $E_{v,v^u}=\Big\{ (v,x) ~|~  x\in V_{vu}\Big\}$ are not present in 
$\widetilde{G^{\prime}}$. 
Therefore, the only way to return from $v^{\prime u}$ to $u$ is to take 
another blue path, which makes the length of the cycle at least $8N-2> 5N+2$.
This implies that a cycle of length $5N+2$ does not exist in this case. \\
\noindent{\it Subcase 1.2:} Let us assume that the cycle does not contain any blue 
edge. Let us assume that the cycle starts at $u \in V(G)$. In this case, the cycle 
starts with an edge $e\in E_{u,u^v}$ for some $v \in N_{G}(u)$. 
Next, it must continue with a red edge. We also observe that if a cycle includes
a red edge from $P_{u^v,v^u}$ then it must include all the red edges of the path
and reaches $v^u$. Again, it must take a 
 $N$ length red edge path as edges in $E_{v,v^u}$ are not present in $\widetilde{G^{\prime}}$. 
 In this way, we observe that a path of length $5N+1$ will end up at
a vertex in the set $V_{vu}$. Since there is no path of length $1$ from a vertex in $V_{vu}$ to $u$, 
we show that such a cycle does not exist. \\
\noindent{\bf Case 2:} If the cycle does not include any original vertex $u\in V(G)$ then 
it also does not include any edge from  $\bigcup\limits_{(u,v)\in E(G)} {E_{u,u^v}} \cup {E_{v,v^u}} \cup E^{\prime}_{u,u^{\prime v}} \cup E^{\prime}_{v,v^{\prime u}} $. Further, we make two subcases based on whether the cycle contains a 
blue edge or not. \\  
\noindent{\it Subcase 2.1:}
Let us assume that the cycle contains a blue edge. In this case, we observe that the length of the cycle is at least $8N-4 > 5N+2$.\\
\noindent{\it Subcase 2.2:} We observe that since the blue edges and the original vertices in 
$V(G)$ are not allowed, the cycle must contain only red edges. In this case we can get 
cycles of even length $2w(u,v)N\neq 5N+2$ only. \\

\par Conversely, suppose $E^{\prime}$ is a solution of the instance $I^{\prime}$. 
First, we show that the set $E^{\prime}$  must contain exactly one of the following four sets 
${E_{u,u^v}} , {E_{v,v^u}} , E^{\prime}_{u,u^{\prime v}}$ or $ E^{\prime}_{v,v^{\prime u}}$
for every $(u,v)\in E(G)$. 
For each edge $(u,v)\in E(G)$, there are $2w(u,v)$ distinct $(u,v)$ paths 
of length $4N$ through the blue edges. We call such paths the paths of type A.
Similarly, for each edge $(u,v)\in E(G)$, there are $2w(u,v)$ distinct $(u,v)$ paths 
of length $N$ through the red edges. We call such paths the paths of type B. 
 We observe that a combination of type A and type B paths form a cycle of length $5N+2$. 
 Therefore, to avoid such a cycle, the solution must destroy all the paths of type A or 
 all the paths of type B. Since the maximum number of edge-disjoint $(u,v)$ path of 
 type A (resp.  B) is $w(u,v)$, the minimum number of edges  whose deletion destroys all  $(u,v)$ paths 
 of type A  (resp.  B) is  $w(u,v)$.   We  must add at least $w(u,v)$ many edges to the solution for each edge 
 $(u,v)$ and 
 since $k=\omega$, it implies that the solution $E^{\prime}$ must include exactly $w(u,v)$ many 
 edges corresponding to each edge $(u,v)\in E(G)$.\\
\noindent{\it Case 1:}
Let us assume that the solution is targeting to destroy all the type B paths. 
In that case, we observe that the solution cannot involve any red edges because if we delete a red edge there are still at least $w(u,v)$ many edge disjoint paths of type B left. 
It implies that solution must contain edges from  ${E_{u,u^v}} \cup {E_{v,v^u}}$. We first observe that we cannot add edges $(u,u_{i}^{v})$ and $(v,v_{i}^{u})$ for any $1\leq i\leq w(u,v)$ in the solution. As otherwise, we will still have $w(u,v)-1$ many edge disjoint $(u,v)$ paths of type B. However, now we are only allowed to delete $w(u,v)-2$ many edges corresponding to edge $(u,v)$.
This is a contradiction as we cannot get rid of all the $5N+2$ length cycles corresponding to the edge $(u,v)$. Without loss 
of generality, we assume that $(u,u_{1}^{v})$ is not part of the solution, that is, we are not deleting $(u,u_{1}^{v})$ from the graph $G'$. 
It forces $(v,v_{1}^{u})$ and $(v,v_{2}^{u})$ to be inside the solution. As $(v,v_{2}^{u})$ is part of the solution implies that $(u,u_{2}^{v})$ is not part of the solution. 
Again, it will force $(v,v_{3}^{u})$ to be part of the solution.
Applying this argument repetitively, we see that $E'$ contains  $E_{v,v^u}$ and since $|{E_{v,v^u}}|=w(u,v)$ implies that no edge from set $E_{u,u^v}$ can be part of the solution. This shows that $E'$  contains either ${E_{u,u^v}}$ or ${E_{v,v^u}}$. \\
\noindent{\it Case 2:}
Let us assume that the solution is targeting to destroy all the type A paths. 
Using the same arguments, we can prove that  $E^{\prime}$  contains either ${E_{u,u^{\prime v}}}$ or ${E_{v,v^{\prime u}}}$.\\

\noindent We define a directed graph $D$ by 
$V(D)=V(G)$ and
\[E(D)=\Big\{ (u,v) ~|~E_{v,v^u} \mbox{ or } E^{\prime}_{v,v^{\prime u}} \subseteq E^{\prime}\Big\}
\bigcup \Big\{ (v,u) ~|~E_{u,u^v} \mbox{ or } E^{\prime}_{u,u^{\prime v}}\subseteq E^{\prime}\Big\}. \]
Suppose there is a vertex $x$ in $D$ for which $w_{\mbox{out}}^x>r$. 
In this case, we observe that $x$ is adjacent to more than $\Delta_w(G) +r$ vertices in graph $\widetilde{G^{\prime}}$. This is a contradiction as vertex $x$ and its neighbours form the
 star graph $S_{\Delta_w(G) +r+1}$, which is a forbidden graph in $I^{\prime}$.

  \section{Hardness of {\sc $\mathcal{F}$-Free Edge Deletion} parameterized by solution size}    
 
 In this section we show that {\sc $\mathcal{F}$-Free Edge Deletion} is W[2]-hard parameterized by the solution size $k$, via 
 a reduction from {\sc Hitting Set}. In the {\sc Hitting Set} problem, we are given a universe $U=\{1,2,\ldots,n\}$, a family 
 $\mathcal{A}$ of sets over $U$, and a positive integer $k$. The objective is to decide whether there is a subset $H\subseteq U$
 of size at most $k$ such that $H$ contains at least one element from each set in $\mathcal{A}$. 
 It is known that  the {\sc Hitting Set} problem is W[2]-hard when parameterized by solution size 
 \cite{marekcygan}. We prove the following theorem:
 
  \begin{theorem}\label{solutiontheorem}\rm
 The {\sc $\mathcal{F}$-Free Edge Deletion} problem is W[2]-hard when parameterized by the solution size $k$,
 the feedback vertex set number or pathwidth of the input graph.
 \end{theorem}

\proof Let  $(U,\mathcal{A},k)$ be an  instance $I$ of the
{\sc Hitting Set} problem and let $U=\{1,2, \ldots, n\}$. We construct an instance $I'=(G,\mathcal{F},k')$ 
of the {\sc $\mathcal{F}$-Free Edge Deletion} problem as follows. We first introduce a {\it central} vertex $v$. 
For every $i \in U$, we attach to this vertex  a cycle $C_{i}$ of length $2i+2$. Note that 
$C_1,C_2, \ldots, C_n$ have only one vertex $v$ in common. We define $G$ as follows $$ V(G) = \bigcup\limits_{i\in U} V(C_{i}) \ \ \text{and} \ \  E(G) = \bigcup\limits_{i\in U} E(C_{i}). $$ 

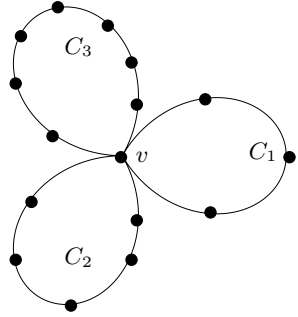
\begin{figure}\small
    \centering
  \begin{tikzpicture}[scale=0.7]
  \centering
   \begin{polaraxis}[grid=none, axis lines=none]
     \addplot[mark=none,domain=0:360,samples=300] { abs(cos(3*x/2))};
   \end{polaraxis}
   
\node[draw, circle, fill=black,inner sep=0pt, minimum size=0.15cm] (v) at (3.4,3.4) [label=right:$v$] {};
\node[draw, circle, fill=black,inner sep=0pt, minimum size=0.15cm](v) at (6.6,3.4) [] {};
\node[draw, circle, fill=black,inner sep=0pt, minimum size=0.15cm](v) at (5,4.5) [] {};
\node[draw, circle, fill=black,inner sep=0pt, minimum size=0.15cm] (v) at (5.1,2.35) [] {};

\node[draw, circle, fill=black,inner sep=0pt, minimum size=0.15cm] (v) at (3.7,2.2) [] {};
\node[draw, circle, fill=black,inner sep=0pt, minimum size=0.15cm] (v) at (1.7,2.55) [] {};
\node[draw, circle, fill=black,inner sep=0pt, minimum size=0.15cm] (v) at (1.4,1.45) [] {};
\node[draw, circle, fill=black,inner sep=0pt, minimum size=0.15cm] (v) at (3.6,1.45) [] {};
\node[draw, circle, fill=black,inner sep=0pt, minimum size=0.15cm] (v) at (2.45,0.58) [] {};

\node[draw, circle, fill=black,inner sep=0pt, minimum size=0.15cm] (v) at (2.1,3.8) [] {};
\node[draw, circle, fill=black,inner sep=0pt, minimum size=0.15cm] (v) at (3.7,4.4) [] {};
\node[draw, circle, fill=black,inner sep=0pt, minimum size=0.15cm] (v) at (3.6,5.2) [] {};
\node[draw, circle, fill=black,inner sep=0pt, minimum size=0.15cm] (v) at (1.4,4.8) [] {};
\node[draw, circle, fill=black,inner sep=0pt, minimum size=0.15cm] (v) at (1.5,5.7) [] {};
\node[draw, circle, fill=black,inner sep=0pt, minimum size=0.15cm] (v) at (3.15,5.9) [] {};
\node[draw, circle, fill=black,inner sep=0pt, minimum size=0.15cm] (v) at (2.2,6.25) [] {};
\node (c1) at (5.5,3.5) [label=right:$C_{1}$] {};
\node (c2) at (2,5.5) [label=right:$C_{3}$] {};
\node (c3) at (2,1.5) [label=right:$C_{2}$] {};
\end{tikzpicture}
 \caption{ The graph $G$ of the {\sc $\mathcal{F}$-Free Edge Deletion} problem  instance constructed in the reduction 
 of Theorem \ref{solutiontheorem} for $n=3$.}
 \label{complementary}
 \end{figure} 
\noindent We observe that the graph $G$ contains a unique cycle $C_{i}$ of length $2i+2$ for each $i\in U$.  
It is clear that $\{v\}$ is a feedback vertex set of $G$. Now, we define a family $\mathcal{F}$ of forbidden 
subgraphs. For every set  $A\in \mathcal{A}$, we add a graph $F_A$ in $\mathcal{F}$, where $F_A$ is defined as follows:
$$ V(F_A) = \bigcup\limits_{i\in A} V(C_i) \ \ \text{and} \ \  E(F_A) = \bigcup\limits_{i\in A} E(C_i). $$ 
We take $k'=k$. Next, we show that $I$ is a yes instance if and only if $I'$ is a yes instance. 
Let $H$ be a solution for the instance $I$. We see that by deleting one arbitrary edge from every 
cycle $C_i$, $i\in H$, we can avoid all the forbidden graphs in $\mathcal{F}$. 
Therefore, we have a solution $E'\subseteq E(G)$ for the instance $I'$ such that $|E'|\leq k'$. 

\par Conversely, suppose  $E'\subseteq E(G)$ with $|E'|\leq k$ is a solution for the instance $I'$.
We see that  $H=\{i ~|~  E(C_{i})\cap E'\neq \emptyset \}$ is a hitting set for the instance $I$. 
We also observe that $|H|\leq k$ as $|H|\leq |E'|$.

% \begin{corollary}\label{twtheorem}\rm
% The {\sc $\mathcal{F}$-Free Edge Deletion} problem is W[2]-hard when parameterized by  the feedback vertex set number or pathwidth of the input graph.
 %\end{corollary}

\begin{corollary}\rm
The {\sc $\mathcal{F}$-Free Edge Deletion} problem is W[2]-hard when parameterized by 
the feedback vertex set number, pathwidth of the input graph and solution size even when restricted to planar, outerplanar, bipartite and planar bipartite graphs.
\end{corollary}

\section{FPT algorithm parameterized by vertex cover number}\label{ndsection}

In this section, we present an FPT algorithm for the  {\sc $\mathcal{T}_{h+1}$-Free Edge Deletion} problem parameterized 
 by the vertex cover number.  A set $C \subseteq V(G)$ is a vertex cover of $G=(V,E)$ if  each edge in $E$ has at least one  endpoint in $C$. In other words, $C$ is a vertex cover of $G$ if and only if $I=V \backslash C$ is an independent set of $G$.  The size of a smallest vertex cover of  $G$ is the vertex cover number of $G$.

 \begin{theorem}\label{theoremnd}\rm
 The {\sc $\mathcal{T}_{h+1}$-Free Edge Deletion} problem is FPT when parameterized by the vertex cover number of the input graph.
 \end{theorem}
 
\proof Without loss of generality we assume that the graph has no isolated vertices. 
Let $S$ be a vertex cover of $G=(V,E)$ of size $k$. We denote by $I$ the independent set 
$V\setminus S$. We partition the independent set $I$ into at most 
$2^{k}$  twin classes $I_{1},I_{2},\dots,I_{2^{k}}$, where some of them can also be empty.
Two vertices $u$ and $v$ are in the same twin class if $N(u)=N(v)$.
Our goal is to minimize the size of $E' \subseteq E(G)$ such that after deleting $E'$ from $G$, 
each connected component  of the resulting graph has at most $h$ vertices.  
First, we  guess the intersection
of $S$ with the connected components in $\widetilde{G}=G \setminus E'$. It is clear that the number of 
guesses is equal to the number of different partitions of the $k$-element set $S$, which is equal to
the Bell number $B_k$. 
For every guess, we will reduce our problem to an  integer linear programming (ILP) where the 
number of variables is a function of the vertex cover number $k$.  Since integer linear 
programming is fixed-parameter tractable when
parameterized by the number of variables, we will conclude that our problem is
fixed-parameter tractable when parameterized by the vertex cover number. 
Let us consider a particular partition
 $P=\{S_{1},S_{2},\dots,S_{\ell}\}$, $\ell\leq k$, of  $S$. For a given partition $P$ of $S$, 
 we call an edge a cross edge 
 if both endpoints of that edge are in $S$ but one endpoint is in $S_{i}$ and other is in $S_{j}$ such that $i \neq j$. 
 We denote the number of cross edges  of partition $P$ by ${\tt cr}(P)$. \\

\noindent{\it ILP Formulation:}  Given a partition $P=\{S_{1},S_{2},\dots,S_{\ell}\}$ of $S$,
let $C_i$ be the component of $\widetilde{G}$ such that $S\cap C_i=S_i$ for $1\leq i\leq \ell$. 
Let $C_{\ell+1}$ be the collection of size one components in $\widetilde{G}$ such that $S\cap C_{\ell+1}=\emptyset$. 
For each $I_i$ and $C_j$, we associate a variable $x_{ij}$ that indicates 
$|I_i\cap C_j|=x_{ij}$, that is,  $x_{ij}$ denotes the number of vertices in twin class $I_{i}$ 
that goes to $C_{j}$.  Because the vertices in $I_i$ have the same neighbourhood, the 
variables $x_{ij}$ determine the  components uniquely and hence 
determine the  required set of edges $E'\subseteq E(G)$.
%We also denote the number of vertices in $I_{j}$ that does not lie in any connected component as 
%$C_{i}$ for all $1\leq i \leq \ell$ by $x_{\ell+1}^{j}$.  
We add the following constraints to ILP. The vertices of each twin class $I_i$ is distributed among
the components $C_1,C_2,\ldots,C_{\ell} \ \text{and} \ C_{\ell+1} $. Thus we have the following constraints:
\begin{align}
    \sum\limits_{j=1}^{\ell+1} x_{ij} =
    |I_i| \ \mbox{ for all } \  1 \leq i \leq 2^{k} 
\end{align}
We want each connected component  $C_j$ in the resulting graph  $\widetilde{G}$ has at most $h$ vertices. 
Thus we have the following constraint: 
\begin{align}
    \sum\limits_{i=1}^{2^{k}} x_{ij} + |S_{i}| \leq h \ \mbox{ for all } \  1 \leq j \leq \ell
\end{align}
Note that every vertex in $I_i$ has the same set of neighbours in $S$.  Thus if a vertex 
$v\in I_i$ goes to $C_j$ then we have to remove all edges between $v$ and $S\setminus S_j$, so that 
$C_1,C_2,\ldots,C_{\ell} \ \text{and} \ C_{\ell+1} $ remains distinct components. Therefore, if $x_{ij}$ vertices of $I_i$ go to 
$C_j$, then we need to remove total $ |N_{S\setminus S_j}(v)| \times x_{ij}$ edges, where $v$ is a vertex in 
$I_i$.
Hence we want to minimize the following objective function: 
\begin{align}
    {\tt cr}(P) + \sum\limits_{i=1}^{2^k}\sum\limits_{j=1}^{\ell+1} {|N_{S\setminus S_j}(v_i)| \times x_{ij}}
\end{align}
where $S_{\ell+1}=\emptyset$, {\tt cr}$(P)$ is the number of cross edges of partition $P$ and 
$v_i$ is a vertex in the twin class $I_i$.

\noindent {\bf Solving the ILP:}
Lenstra \cite{lenstra} showed that the feasibility version of {\sc $p$-ILP} is FPT with 
running time doubly exponential in $p$, where $p$ is the number of variables. 
Later, Kannan \cite{kannan} proved an algorithm for {\sc $p$-ILP} running in time $p^{O(p)}$.
In our algorithm, we need the optimization version of {\sc $p$-ILP} rather than 
the feasibility version. We state the minimization version of {\sc $p$-ILP}
as presented by Fellows et. al. \cite{fellows}. \\

\noindent {\sc $p$-Variable Integer Linear Programming Optimization ($p$-Opt-ILP)}: Let matrices $A\in \ Z^{m\times p}$, $b\in \ Z^{p\times 1}$ and 
$c\in \ Z^{1\times p}$ be given. We want to find a vector $ x\in \ Z ^{p\times 1}$ that minimizes the objective function $c\cdot x$ and satisfies the $m$ 
inequalities, that is, $A\cdot x\geq b$.  The number of variables $p$ is the parameter. 
Then they showed the following:

\begin{lemma}\rm \label{ilp}\cite{fellows}
{\sc $p$-Opt-ILP} can be solved using $O(p^{2.5p+o(p)}\cdot L \cdot log(MN))$ arithmetic operations and space polynomial in $L$. 
Here $L$ is the number of bits in the input, $N$ is the maximum absolute value any variable can take, and $M$ is an upper bound on 
the absolute value of the minimum taken by the objective function.
\end{lemma}

In the formulation for {\sc $\mathcal{T}_{h+1}$-Free Edge Deletion} problem, we have at most $2^{k}(k+1)$ 
variables. The value of objective function is bounded by $n^2$ and the value of any variable 
in the integer linear programming is  bounded by $n$. The constraints can be represented using at most
$O(2^k \log{n})$ bits. Lemma \ref{ilp} implies that we can solve the problem with the guess $P$ in FPT time. 
There are at most $B_k$ choices for $P$, and the ILP formula for a guess can be solved in FPT time. Thus 
Theorem \ref{theoremnd} holds.

\section{FPT algorithm parameterized by combined parameters $k$ and $h$}

In this section we give a kernelization algorithm for the {\sc $\mathcal{T}_{h+1}$-Free Edge Deletion} problem
based on a reduction rule. For a given instance $(G,k,h)$ of the {\sc $\mathcal{T}_{h+1}$-Free Edge Deletion} problem
if $G$ has a component of size at most $h$, then its removal does not change the solution. 
This shows that the following rule is safe.

\noindent{\bf Reduction 1:} If $G$ contains a component $C$ of size at most $h$, then delete $C$ from $G$, 
the new instance is $(G-C, k, h)$.\\

\noindent This leads to the following lemma. 

\begin{lemma}\label{twtheorem1}
If $(G,k,h)$ is a yes-instance and Reduction rule 1 is not applicable to $G$, then $|V(G)|\leq 2kh$ and 
$|E(G)|\leq 2kh^2+k$.
 \end{lemma}
 
 \proof Because we cannot apply Reduction rule 1, $G$ has no components of size at most $h$. 
 Since $(G,k,h)$ is a yes-instance, there is a subset $E'\subseteq E(G)$ such that $|E'|=k$ and 
 every component of $G\setminus E'$ has at most $h$ vertices. If we put back the $k$ edges, as one edge
 can join two components, $k$ edges of $E'$ can join at most $2k$ components. This implies that the number of connected 
 components
 of $G$ is bounded by $2k$. As there are at most $2k$ components, we get $|V(G)|\leq 2kh$. Since each component can
 have at most $h^2$ edges and there are $2k$ components, we get $|E(G)|\leq 2kh^2+k$.\\
 
 \noindent  Finally, we remark that the Reduction rule 1 is applicable in $O(V+E)$ time. Thus we obtain the following theorem
 
\begin{theorem}\label{twtheorem2}
 The {\sc $\mathcal{T}_{h+1}$-Free Edge Deletion} problem  admits a kernel with $O(hk)$ vertices and 
 $O(h^2k)$ edges.  
 \end{theorem}
 
 \section{Conclusions and Open Problems} The main contributions in this paper are that  the {\sc $\mathcal{F}$-Free Edge Deletion} problem is W[1]-hard
when parameterized by treewidth; it is W[2]-hard when parameterized by the solution size, pathwidth or feedback vertex set number; 
the {\sc $\mathcal{T}_{h+1}$-Free Edge Deletion} problem is FPT when parameterized by vertex cover number; and it is 
FPT when parameterized by combined parameters $k$ and $h$.  
We list some nice problems emerge from the results here: 
does {\sc $\mathcal{T}_{h+1}$-Free Edge Deletion} admit a polynomial kernel in vertex cover? Also, noting that 
the problem is FPT in vertex cover, it would be interesting to consider the parameterized complexity with respect to twin cover. 
The modular width parameter also appears to be a natural parameter to consider here.  
The parameterized complexity of the problem 
remains unsettle  when parameterized by other important 
structural graph parameters like clique-width.   As mentioned in \cite{kitty}, one problem of practical relevance to 
epidemiology would be the complexity of the problems on planar graph; this would be relevant for considering the spread of 
a disease based on the geographic location of animal holdings.

\bibliographystyle{abbrv}
\bibliography{bibliography}

\begin{thebibliography}{10}

\bibitem{CAI1996171}
L.~Cai.
\newblock Fixed-parameter tractability of graph modification problems for
  hereditary properties.
\newblock {\em Information Processing Letters}, 58(4):171 -- 176, 1996.

\bibitem{marekcygan}
M.~Cygan, F.~V. Fomin, L.~Kowalik, D.~Lokshtanov, D.~Marx, M.~Pilipczuk,
  M.~Pilipczuk, and S.~Saurabh.
\newblock {\em Parameterized Algorithms}.
\newblock Springer, 2015.

\bibitem{Downey}
R.~G. Downey and M.~R. Fellows.
\newblock {\em Parameterized Complexity}.
\newblock Springer, 2012.

\bibitem{kitty}
J.~Enright and K.~Meeks.
\newblock Deleting edges to restrict the size of an epidemic: A new application
  for treewidth.
\newblock {\em Algorithmica}, 80(6):1857--1889, 2018.

\bibitem{fellows}
M.~R. Fellows, D.~Lokshtanov, N.~Misra, F.~A. Rosamond, and S.~Saurabh.
\newblock Graph layout problems parameterized by vertex cover.
\newblock In S.-H. Hong, H.~Nagamochi, and T.~Fukunaga, editors, {\em
  Algorithms and Computation}, pages 294--305, Berlin, Heidelberg, 2008.
  Springer Berlin Heidelberg.

\bibitem{FUJITO1998213}
T.~Fujito.
\newblock A unified approximation algorithm for node-deletion problems.
\newblock {\em Discrete Applied Mathematics}, 86(2):213 -- 231, 1998.

\bibitem{Esha}
E.~Ghosh, S.~Kolay, M.~Kumar, P.~Misra, F.~Panolan, A.~Rai, and M.~S.
  Ramanujan.
\newblock Faster parameterized algorithms for deletion to split graphs.
\newblock {\em Algorithmica}, 71(4):989--1006, 2015.

\bibitem{Gibbens729}
J.~C. Gibbens, J.~W. Wilesmith, C.~E. Sharpe, L.~M. Mansley, E.~Michalopoulou,
  J.~B.~M. Ryan, and M.~Hudson.
\newblock Descriptive epidemiology of the 2001 foot-and-mouth disease epidemic
  in great britain: the first five months.
\newblock {\em Veterinary Record}, 149(24):729--743, 2001.

\bibitem{Guo}
J.~Guo.
\newblock Problem kernels for np-complete edge deletion problems: Split and
  related graphs.
\newblock In {\em Proceedings of the 18th International Conference on
  Algorithms and Computation}, ISAAC'07, pages 915--926, Berlin, Heidelberg,
  2007. Springer-Verlag.

\bibitem{kannan}
R.~Kannan.
\newblock Minkowski's convex body theorem and integer programming.
\newblock {\em Mathematics of Operations Research}, 12(3):415--440, 1987.

\bibitem{Danon}
B.~Kerr, L.~Danon, A.~P. Ford, T.~House, C.~P. Jewell, M.~J. Keeling, G.~O.
  Roberts, J.~V. Ross, and M.~C. Vernon.
\newblock Networks and the epidemiology of infectious disease.
\newblock {\em Interdisciplinary Perspectives on Infectious Diseases},
  2011:284909, 2011.

\bibitem{lenstra}
H.~W. Lenstra.
\newblock Integer programming with a fixed number of variables.
\newblock {\em Mathematics of Operations Research}, 8(4):538--548, 1983.

\bibitem{Lund}
C.~Lund and M.~Yannakakis.
\newblock On the hardness of approximating minimization problems.
\newblock {\em J. ACM}, 41(5):960--981, Sept. 1994.

\bibitem{Mansley43}
L.~M. Mansley, P.~J. Dunlop, S.~M. Whiteside, and R.~G.~H. Smith.
\newblock Early dissemination of foot-and-mouth disease virus through sheep
  marketing in february 2001.
\newblock {\em Veterinary Record}, 153(2):43--50, 2003.

\bibitem{NATANZON2001109}
A.~Natanzon, R.~Shamir, and R.~Sharan.
\newblock Complexity classification of some edge modification problems.
\newblock {\em Discrete Applied Mathematics}, 113(1):109 -- 128, 2001.

\bibitem{Neil}
N.~Robertson and P.~Seymour.
\newblock Graph minors. iii. planar tree-width.
\newblock {\em Journal of Combinatorial Theory, Series B}, 36(1):49 -- 64,
  1984.

\bibitem{DBLP:journals/corr/abs-1107-1177}
S.~Szeider.
\newblock Not so easy problems for tree decomposable graphs.
\newblock {\em CoRR}, abs/1107.1177, 2011.

\bibitem{WATANABE198363}
T.~Watanabe, T.~Ae, and A.~Nakamura.
\newblock On the np-hardness of edge-deletion and -contraction problems.
\newblock {\em Discrete Applied Mathematics}, 6(1):63 -- 78, 1983.

\bibitem{10.1145/800133.804355}
M.~Yannakakis.
\newblock Node-and edge-deletion np-complete problems.
\newblock In {\em Proceedings of the Tenth Annual ACM Symposium on Theory of
  Computing}, STOC '78, pages 253--264, New York, NY, USA, 1978. Association
  for Computing Machinery.

\end{thebibliography}
\newpage

\appendix

\section{Preliminaries}    
Unless otherwise stated all graphs are simple, undirected, and loopless. 
For graph $G=(V,E)$, $V=V(G)$ is the vertex set of $G$, and $E=E(G)$ the
edge set of $G$. We now recall  some graph parameters used in this paper. The graph
parameters we explicitly use in this paper are feedback vertex set  and treewidth. 
\begin{definition}\rm 
A feedback vertex set in an undirected graph $G$ is a subset of vertices whose removal results in an acyclic graph.
The minimum size of a feedback vertex set in $G$ is the {\it feedback vertex set number} of $G$, denoted by 
{\tt fvc}$(G)$.
\end{definition}
We now review the concept of a tree decomposition, introduced by Robertson and Seymour in \cite{Neil}.
\begin{definition}\rm  A {\it tree decomposition} of a graph $G$ is a pair $(T,\{X_{t}\}_{t\in V(T)})$,
where $T$ is a tree and each node $t$ of the tree $T$ is assigned a vertex subset  
$X_{t} \subseteq V(G)$, called a bag,  such that  the following conditions are satisfied:
		\begin{enumerate}
			\item Every vertex of $G$ is  in at least one bag.
			\item For every edge $uv \in E(G)$, there exists a node $t\in T$ 
			such that bag $X_t$ contains  both $u$ and $v$.
			\item For every $u \in V(G)$,  the set $\{t\in V(T)~|~u\in X_t\}$ induces a connected subtree of $T$. 
		\end{enumerate}
	\end{definition}
	
%\noindent It is important to note that a graph may have several different tree decomposition.
%	Similarly, the same tree decomposition can be valid for several different graphs. 
%	Every graph has a trivial tree decomposition for which $T$ has only one vertex including all
%	of $V$. However, this is not effective for the purpose of solving problems. 
	
\begin{definition} \rm The {\it width} of a tree decomposition is defined as $width(T)=max_{t \in V(T)}|X_{t}|-1$ and the treewidth $tw(G)$ of a graph $G$  is the  minimum width among all possible tree decomposition of $G$.
\end{definition}

\end{document}